\documentstyle[12pt,aps,prc,epsfig,preprint]{revtex}
\tightenlines
\begin{document}

\begin{titlepage}
\title{\vspace*{10mm}\bf On the determination of the pion effective
mass in nuclei from pionic atoms}
\vspace{6pt}

\author{ E.~Friedman and A.~Gal \\
{\it Racah Institute of Physics, The Hebrew University, Jerusalem 91904,
Israel\\}}

\vspace{4pt}
\maketitle

\begin{abstract}
The binding energies of the deeply bound 1s and 2p  states
in pionic atoms of $^{207}$Pb, recently established experimentally 
in the $^{208}$Pb(d,$^3$He) reaction, have been used by several groups to
derive the pion effective mass in nuclear matter. We show that these
binding energies are fully consistent with `normal' pionic atoms and that
the real part of the pion-nucleus potential at the center of $^{207}$Pb
is 28$\pm$3 MeV and not 20 MeV as suggested previously.
\newline \newline
$PACS$: 36.10.Gv; 14.40.Aq
\newline
Corresponding author: E. Friedman,\newline
Tel: +972 2 658 4667, 
Fax: +972 2 658 6347, \newline
E mail: elifried@vms.huji.ac.il
\end{abstract}
\centerline{\today}
\end{titlepage}

Information on the strong interaction at zero energy between a negatively
charged hadron and a nucleus may be obtained from the observation of 
level shifts and widths in hadronic atoms. Such  levels are populated
via an atomic cascade process of the hadron  where 
the experimentally
observed X-ray spectra are terminated at a level for which the radiation 
yield becomes smaller than the nuclear absorption. Only for  light
pionic atoms is the 1s level observed, otherwise the spectrum terminates
at the 2p, 3d or higher levels. 
Simple extrapolations led to the expectation that
1s and 2p states in heavy pionic atoms would be quite broad due to the
nuclear absorption. 
The first  to show that 1s and 2p states in heavy pionic atoms
 are so narrow  as to make them well defined, were
Friedman and Soff \cite{FSo85}. They calculated shifts and widths
for pionic atom states well beyond the experimentally reachable region
and showed that 
the atomic wavefunctions of these deeply bound states are pushed out of
the nucleus by the repulsive $s$-wave part of the potential
such that their overlap with the nucleus and with the imaginary
part of the potential becomes very small. 
Similar conclusions about the small widths expected for deeply
bound pionic atom states were reached by Toki and Yamazaki
\cite{TYa88,THY89}, who  also considered methods, other than
radiative processes, to populate such states. The pionic 1s and 2p states
in $^{207}$Pb were observed recently by Yamazaki et al. \cite{Yam96,Yam97}
in the $^{208}$Pb(d,$^3$He) reaction, yielding for the binding energies
the values \cite{Yam97}

\begin{equation}\label{Bexp}
{\rm B}_{{\rm 1s}}=7.1\pm0.2 ~{\rm MeV}\quad\quad {\rm B}_{{\rm 2p}}
=5.31\pm0.09~ {\rm MeV}.
\end{equation}

\noindent
 The availability of such results raises the question of the consistency of
these deeply bound states with the `normal' pionic atom states, within the
commonly accepted pion-nucleus interaction model. It also focuses 
attention \cite{WBW97} on the pion effective mass in the nuclear
medium.  These two points are the topics discussed in this Letter.

The pion-nucleus potential at zero energy is traditionally written
\cite{EEr66} in the form
\begin{eqnarray} \label{EE1}
2\mu V_{opt}(r) = q(r) + \vec \nabla \cdot \alpha(r) \vec \nabla 
\end{eqnarray}

\noindent
with the $s$-wave part given by 

\begin{eqnarray} \label{EE1s}
q(r) & = & -4\pi(1+\frac{\mu}{M})\{b_0[\rho_n(r)+\rho_p(r)]
  +b_1[\rho_n(r)-\rho_p(r)] \} \nonumber \\
 & &  -4\pi(1+\frac{\mu}{2M})4B_0\rho_n(r)\rho_p(r).
\end{eqnarray}

\noindent
In this expression $\rho_n$ and $\rho_p$ are the neutron and proton density
distributions normalized to the number of neutrons $N$ and number
of protons $Z$, respectively, $\mu$ is the pion-nucleus reduced mass
and $M$ is the mass of the nucleon. The real coefficients $b_0$ and
$b_1$, according to the low density limit \cite{DHL71}, are expressed
by the $\pi^-$p elastic scattering and charge exchange scattering lengths,
which have been determined recently from pionic hydrogen \cite{Sig96}:

\begin{equation} \label{b0b1}
b_0=-0.0077\pm0.0072 ~m_\pi^{-1},
 \quad \quad  b_1=-0.0962\pm0.0071 ~m_\pi^{-1}.
\end{equation}

\noindent
These values, as summarized in Ref.\cite{WBW97}, agree very well with
those calculated by chiral perturbation theory.
The parameter $B_0$ is obtained phenomenologically   from fits
to pionic atom data and its
 imaginary part  represents $s$-wave  absorption on two
nucleons, which is dominated by absorption on a neutron-proton pair.
A  real part for $B_0$ cannot be excluded and indeed is found
to be required by fits to the data, even when $b_0$ and $b_1$ are treated
as free parameters. 
This real part is referred to as
the `missing' $s$-wave repulsion  
because it turns out to be repulsive and substantially
larger than the imaginary
part   (Ref. \cite{BFG97} and references therein),
 contrary to expectations. 
A correlation between the parameters $b_0$ and Re$B_0$ was noted long
ago \cite{SMa83} on the basis of analyses of older and much more
restricted data sets, suggesting that $b_0$ and Re$B_0$ can be lumped
together. However,  recent analyses
of considerably more extended data bases \cite{BFG97,KLT90} find each of the
parameters  to be reasonably well determined (see table 3 
of Ref. \cite{BFG97}).
Finally, when discussing the real part of the $s$-wave potential one has to
realize that the isoscalar coefficient $b_0$ is exceptionally small.
Therefore an explicit second order term is often included in the
isoscalar part, with $b_0$  replaced by 

\begin{equation} \label{b0b}
\overline{b}_0 = b_0 - \frac{3}{2\pi}(b_0^2+2b_1^2)k_F
\end{equation}

\noindent
where $k_F$ is the Fermi momentum taken either as a constant, or
calculated for the local nuclear density. In the present work we adopt 
this additional term with the latter prescription.

The data base for normal pionic atoms used in the present work 
 contains 54 data points for 1s to 4f states  covering
the range from oxygen to uranium. It was shown in \cite{BFG97} that it
leads to essentially the same results as the very extended data base of
Konijn et al. \cite{KLT90} which contains 140 points. 
As a starting point we note that an unconstrained fit to the data
for `normal' states leads to 
$\chi^2$/N, the $\chi^2$ per point, of 2.0, a value which is used
 as a reference
to the quality of subsequent fits. The calculated 
binding energies for the 1s and 2p
states in $^{207}$Pb are 6.77 and 5.10 MeV, respectively, compared
to the experimental values Eq.(\ref{Bexp}),
which means a somewhat inferior agreement between calculation and experiment
for the deeply bound states compared to the normal states. The value of the
$s$-wave potential at the center of the $^{207}$Pb nucleus, $V_S$, 
is found to be 29 MeV 
(repulsive). Next we adopted a more  constrained approach where the 
values of $b_0$ and $b_1$, and of the linear terms in the $p$-wave $\alpha(r)$ 
part of the potential
(not discussed in the present work), were held fixed at 
the corresponding free $\pi$N
values in order to respect the low density limit. 
This fit leads to $\chi^2$ per
point of 2.9 and the predicted values of the 1s and 2p binding energies
are then 6.84 and 5.14 MeV, respectively. The $\chi^2$ per point for these two
states is now 2.6, thus suggesting full consistency with the normal states.
The $s$-wave potential at the center of $^{207}$Pb is now $V_S$=27.0 MeV,
with Re$B_0$=$-$0.062$\pm$0.006 m$_\pi^{-4}$, 
Im$B_0$=0.056$\pm$0.003~m$_\pi^{-4}$, thus demonstrating the importance of
the  Re$B_0$ term in addition to the free $\pi$N $s$-wave
interaction terms, a point which
 is at variance with conclusions of Ref. \cite{WBW97}.
Furthermore, the resulting value for $V_S$ disagrees with that found 
by considering {\it only} the deeply bound states \cite{Yam97,WBW97}
without checking for consistency with the normal data. 
Indeed by setting Re$B_0$=0
and fitting $b_0$ and $b_1$ to the experimental 1s and 2p binding energies
(and using the standard $p$-wave potential \cite{BFG97}),  we find
$V_S$=16.9 MeV in agreement with \cite{Yam97,WBW97}. However, this potential
results in $\chi^2$/N=54 for the normal states, which is totally unacceptable.
 We therefore proceed with a more systematic study of the connection between
Re$B_0$ and $V_S$.

\begin{figure}
\epsfig{file=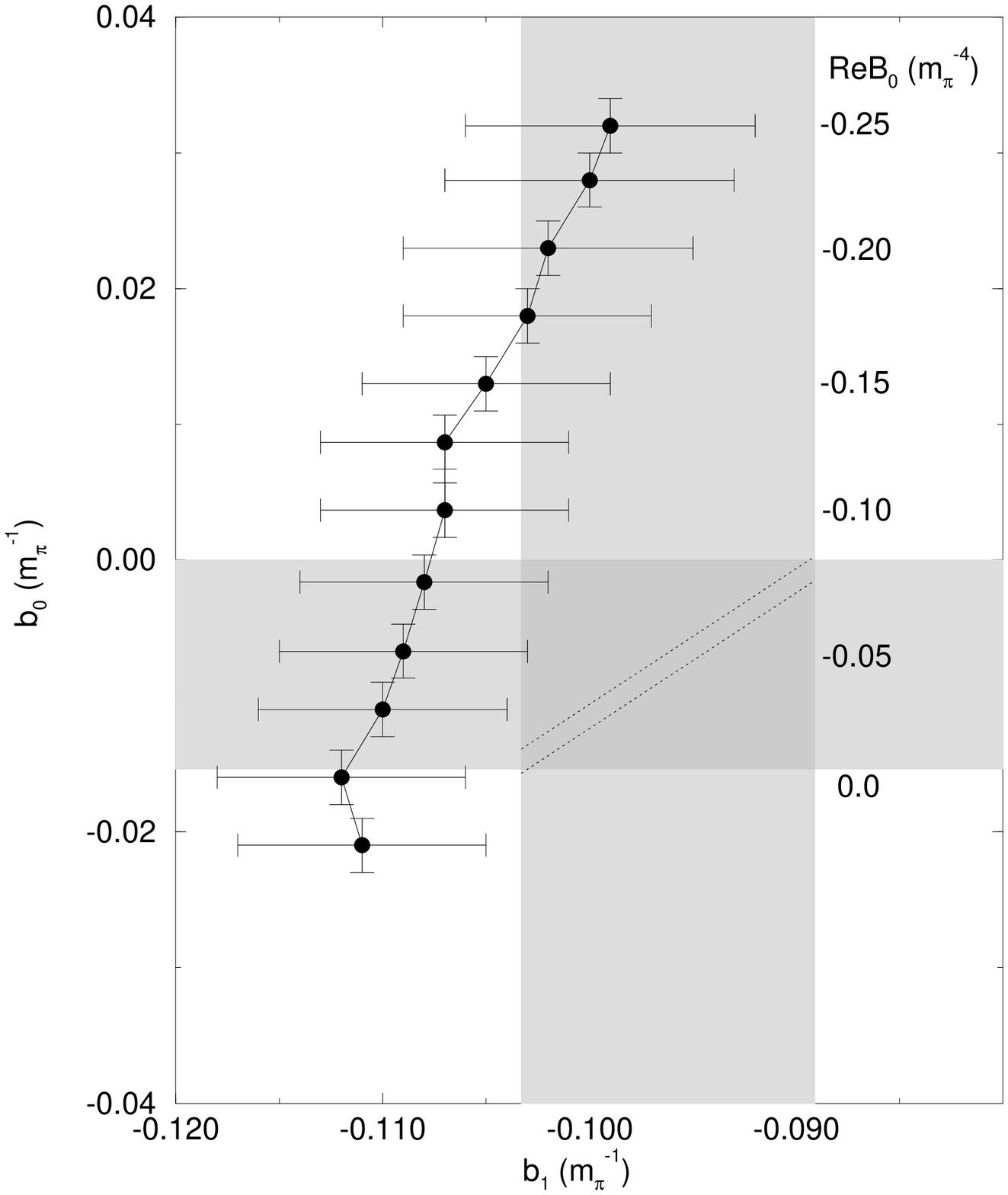,height=160mm,width=135mm,
bbllx=17,bblly=86,bburx=524,bbury=691}
\caption{Values of $b_0$ and $b_1$ obtained from $\chi^2$ fits to normal
pionic atom data are denoted by solid dots.
 Values of Re$B_0$ were held fixed during the fits and are
listed in the figure (see text for details).}
\label{fig:b0b1}
\end{figure}

\begin{figure}
\epsfig{file=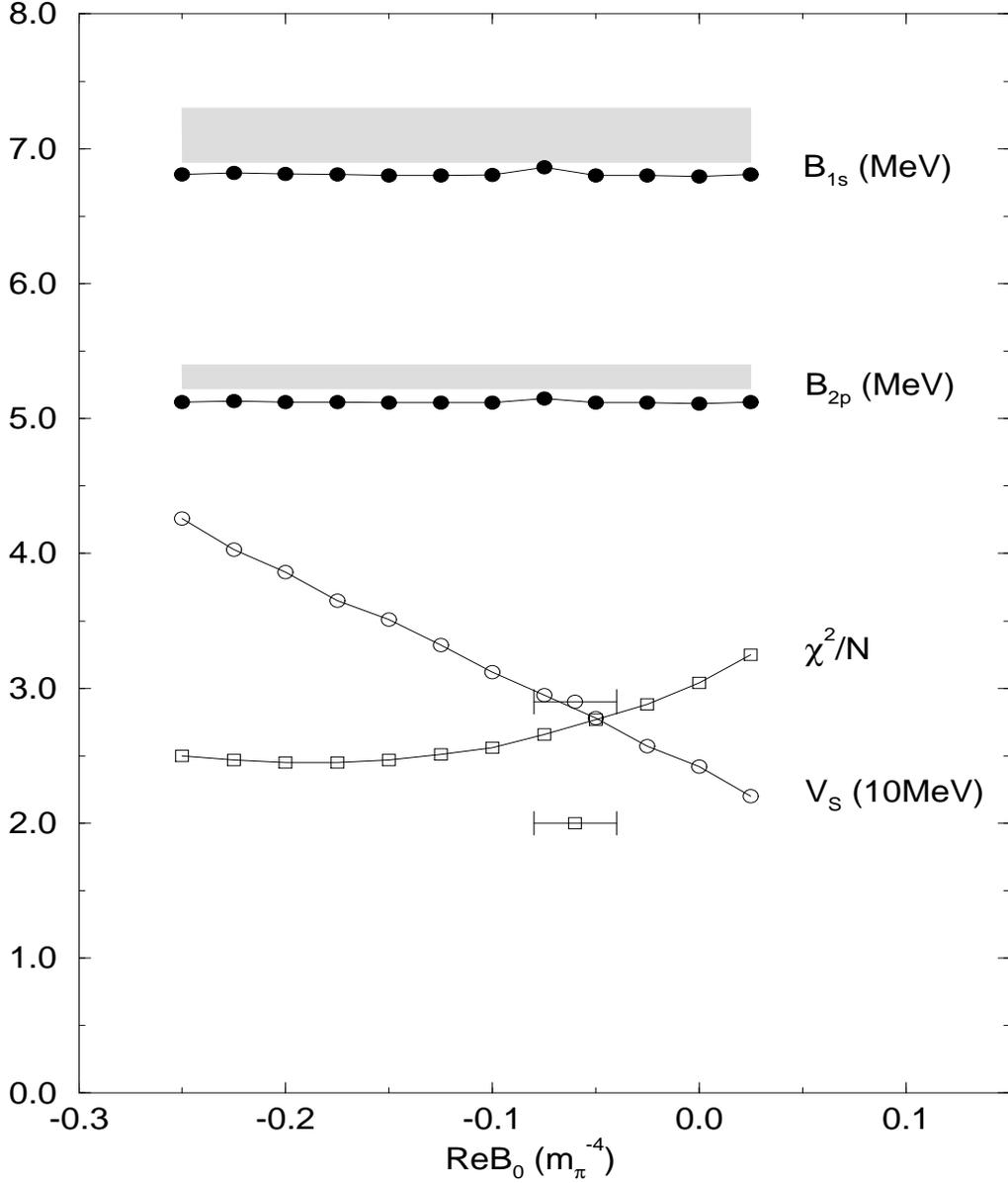,height=160mm,width=135mm,
bbllx=74,bblly=88,bburx=508,bbury=670}
\caption{Values of $\chi^2$/N, of $V_S$ and of the the calculated
binding energies of pionic 1s and 2p states in $^{207}$Pb as functions
of Re$B_0$. The dots correspond to the points on Fig. \ref{fig:b0b1}.
The shaded bands are the experimental results for the binding energies.
The two horizontal error bars correspond to the unconstrained best fit.}
\label{fig:allB}
\end{figure}

The role of the parameter Re$B_0$ can be assessed by gridding on its value
while performing fits to the data, 
 varying the other parameters, {\it including} $b_0$ {\it and} $b_1$.
Figure \ref{fig:b0b1} shows the best fit
values of $b_0$ and $b_1$ together with
their uncertainties, along  with the fixed values of
Re$B_0$ listed on the right hand side. 
Also shown as shaded bands are the 
experimental values \cite{Sig96} of $b_0$ and $b_1$ as obtained from 
pionic hydrogen. The correlation between these two parameters for the free
pion-nucleon interaction is determined to  very high precision \cite{Sig96},
 as is
depicted by the two dashed lines within the central shaded area.
Figure \ref{fig:allB} shows the corresponding values of $\chi^2$/N,
of $V_S$ and of the calculated binding energies of the 1s and 2p
states in $^{207}$Pb. The experimental  values of these two binding
energies are shown as  shaded areas. The two horizontal error bars
show the best $\chi^2$/N and the corresponding value of $V_S$
for the unconstrained fit
mentioned above. The dots in this figure correspond to the dots
in Fig. \ref{fig:b0b1}. The calculated binding energies for the 1s and 2p
states hardly vary, demonstarting the $b_0$-Re$B_0$ ambiguity noted
by Toki et al. \cite{THY89}.
 The differences between calculated and
experimental results for the 1s and 2p binding energies
show a systematic trend. However, this difference amounts to $\chi^2$/N=2.5
for these two states, which is about the same as obtained
for the 54 data points of the normal pionic atom states.
 It is therefore concluded
that the two new results for the deeply bound pionic states
\cite{Yam97} are fully consistent with the normal states and do not seem to 
convey any new information about the pion nucleus interaction.

The values of $V_S$, which are taken as representative of the real
$s$-wave potential in nuclear matter, deserve some discussion. It is
clear from Fig. \ref{fig:allB} that fits to pionic atom data do not
determine  $V_S$ at all and 
 its value depends sensitively on the assumptions
made about  Re$B_0$. A similar conclusion is obtained when fits are made to the
1s and 2p states only.
Nevertheless, it can be shown by performing a `notch test' \cite{BFG97}
that once the coefficients in $V_{opt}$ (Eq.(\ref{EE1})) are fixed, the values
of the potential well inside the nucleus are determined within a few percent. 
If, for example, one assumes that Re$B_0$=0 then $V_S$ 
becomes close to 24 MeV.  However, this 
would lead to a deterioration in the fits (Fig. \ref{fig:allB})
to normal pionic atom data.
 Imposing the free values on $b_0$ and $b_1$ 
whilst keeping Re$B_0$=0,
reduces further the value of $V_S$ to 17 MeV 
while causing $\chi^2$/N to become 19, which is unacceptably large
in view of the smaller values shown in Fig. \ref{fig:allB}. Requiring
$\chi^2$/N to be around 3 for normal pionic atom states and requiring
values for $b_0$ and $b_1$ which are removed from the free $\pi$N
values by no more than one standard deviation, the two figures constrain
Re$B_0$ to be between $-$0.02 and $-$0.08 m$_\pi^{-4}$. This leads to
the value of the real part of the repulsive $s$-wave potential inside
$^{207}$Pb to be  28$\pm$3 MeV. It implies
a value of $m_\pi(\rho)=$170.4$\pm$3.6 MeV, where the pion effective
mass at density $\rho$ is given by

\begin{equation} \label{effm}
m_{\pi}^2(\rho)=m_{\pi}^2+2m_{\pi}(\rho)V_S(\rho), \quad\quad m_{\pi}(0)=m_\pi.
\end{equation}

\noindent
For symmetric nuclear matter ($\rho_0$=0.17 fm$^{-3}$)
where $b_1$ is ineffective, one then has
$m_{\pi}(\rho_0)=$167 MeV.

A possible point of concern is the poorly known density distributions for
neutrons which enter the optical potentials. For $N=Z$ nuclei it is obvious
to use the same distribution for the neutrons as for the protons. For $N>Z$
nuclei we followed the general procedure outlined in \cite{BFG97} using
for  neutrons distributions with slightly larger rms radii than for the
protons. For $^{207}$Pb the rms radius for the neutrons was chosen as 0.19 fm
larger than the corresponding value for the protons, as suggested by averaging
the various results for $^{208}$Pb summarised in \cite{BFG89}. However, recent
calculations based on relativistic mean field theory \cite{WNP98}
suggest a value of 0.26 fm
for the difference. The sensitivities of the various calculated values for
$^{207}$Pb for a change of 0.1 fm in that difference are $-$95 keV for
B$_{{\rm 1s}}$, $-$55 keV for B$_{{\rm 2p}}$ and $-$2.7 MeV for $V_S$. The value
of $m_{\pi}(\rho_0)$, the effective mass in symmetric nuclear matter, depends
only very marginally on the assumed radii for the neutron distributions.

A comment on the widths of the deeply bound states is in order.
Experimentally \cite{Yam96,Yam97} only an upper limit of 0.8 MeV could be
placed on the width of the 2p state. All our potentials that produce
acceptable fits to normal pionic atoms predict a width of 0.31-0.33 MeV
for the 2p state, and a width of 0.45-0.50 MeV for the 1s state. However,
if one sets Re$B_0$=0 and imposing on $b_0$ and $b_1$ the 
free $\pi$N value, then
the widths of the 2p and 1s states become 0.44 and 0.8  MeV, respectively.
Much improved experimental accuracies will be required in the determination
of widths of deeply bound states if these are to be useful as a source
of information on the $\pi$-nucleus interaction.

In conclusion, fits to normal pionic atom data show that a 
real part of the two-nucleon absorption term of the $s$-wave part of the
potential is required and is
well determined by the data, once the low density limit is imposed
on the $s$-wave part of $V_{opt}$ (Eq. (\ref{EE1})),
 thus demonstrating that the problem of `missing' $s$-wave repulsion
persists \cite{BFG97}.
Imposing the low density limit on the $s$-wave part of the potential, its
value in the interior of the $^{207}$Pb nucleus is found to be 28$\pm$3 MeV,
leading to a pion effective mass in {\it symmetric} nuclear matter 
of 167$\pm$3.5 MeV.
It is also shown that any {\it good}
fit to normal pionic atom data, which 
approximately respects the low density limit,
leads to calculated binding energies for the deeply bound 1s and 2p
states in $^{207}$Pb that agree with experiment at the same level as for
normal states and to $V_S$ in the range specified above.
 One might suspect that the same mechanism  which
causes the deeply bound states to be narrow also masks the deep
interior of nuclei where new effects could possibly be observed.

\vspace{15mm}

We wish to thank C.J. Batty for useful comments.
This research was partially supported by the Israel Science Foundation.

\end{document}